\documentclass[journal]{IEEEtran}
\ifCLASSINFOpdf
\else
\fi

\usepackage{amsmath}
\usepackage{graphicx}
\usepackage[dvipsnames]{xcolor}
\begin{document}
%
\title{Hopf Physical Reservoir Computer for Reconfigurable Sound Recognition}
%
%
%

\author{Md Raf E Ul Shougat$^{1}$, XiaoFu Li$^{2}$, Siyao Shao$^{3}$, Kathleen McGarvey$^{4}$, and Edmon Perkins$^{5*}$

\thanks{$^{1}$Md Raf E Ul Shougat is a  PhD candidate at North Carolina State University, Raleigh, NC 27695, USA {\tt\small mdrafeulshougat@ncsu.edu}.}
\thanks{$^{2}$XiaoFu Li is a researcher at LAB2701, Atwood, OK 74827, USA {\tt\small xfxyyh0812@gmail.com}.}
\thanks{$^{3}$Siyao Shao is a co-founder of \textit{echosonic}, Montreal, Quebec, Canada {\tt\small siyao@echosonic.ca}.}%
\thanks{$^{4}$Kathleen McGarvey is a venture development specialist at TandemLaunch {\tt\small kathleen.mcgarvey@tandemlaunch.com}.}%
\thanks{$^{5*}$Edmon Perkins is the owner of LAB2701, Atwood, OK 74827, USA {\tt\small edmon@lab2701.com}.}
}

%
%

\markboth{IEEE Internet of Things Journal}%
{Shougat \MakeLowercase{\textit{et al.}}: Hopf PRC for Reconfigurable Sound Recognition}
%



\maketitle

\begin{abstract}
The Hopf oscillator is a nonlinear oscillator that exhibits limit cycle motion. This reservoir computer utilizes the vibratory nature of the oscillator, which makes it an ideal candidate for reconfigurable sound recognition tasks. In this paper, the capabilities of the Hopf reservoir computer performing sound recognition are systematically demonstrated. This work shows that the Hopf reservoir computer can offer superior sound recognition accuracy compared to legacy approaches (e.g., a Mel spectrum + machine learning approach). More importantly, the Hopf reservoir computer operating as a sound recognition system does not require audio preprocessing and has a very simple setup while still offering a high degree of reconfigurability. These features pave the way of applying physical reservoir computing for sound recognition in low power edge devices.
\end{abstract}


%
\IEEEpeerreviewmaketitle

\section{Introduction}

\IEEEPARstart{T}{here} are ubiquitous methods of audio signal classification, particularly for speech recognition \cite{lee2021biosignal, karmakar2021thank}. However, machine learning suffers several drawbacks that hinder its wide dissemination on the Internet of Things (IoT) \cite{filho2022systematic}. First, machine learning, especially deep neural networks (DNNs), rely on the cloud infrastructure to conduct massive computation for both model training and inference. State-of-the-art (SOTA) deep learning models, such as GPT-3, can have over 175 billion parameters and training requirements of 3.14 $\times$ $10^{23}$ FLOPS (floating operations per second) \cite{li2020openai, patterson2022carbon}. The training of the SOTA speech transcription model, Whisper, used a word library that had as many words as one person would continuously speaks for 77 years \cite{radfordrobust}. None of these mentioned technical requirements could be fulfilled by any edge devices for IoT; thus, the cloud infrastructure is a necessity for DNN tasks. Second, reliance of cloud computing for machine learning poses great security and privacy risks. Over 60\% of previous security breaches happened during the raw data communication between the cloud and the edge for machine learning \cite{adversa}. Further, each breach carries an average \$4.24 million loss, and this number is continuously growing \cite{ibm}. The privacy concern causes distrust among smart device users and drives the abandonment of smart devices \cite{garg2018open, deep2022survey}. Third, the environmental impact of implementing DNN through a cloud infrastructure is often overlooked but cannot be neglected. Training a transformer model with 213 million parameters will generate carbon dioxide emissions equaling four times of a US manufacturer's vehicle over its whole lifespan \cite{hao2019training}. Therefore, the next generation of smart IoT devices needs to possess sufficient computational power to operate machine learning or even deep learning on the edge. 

Among efforts to bring machine learning to edge devices, reservoir computing, especially physical reservoir computing, has generated early success over the last two decades. Originating from the concepts of liquid state machines and echo state networks, researchers demonstrated that the sound-induced ripples on the surface of a bucket of water could be used to conduct audio signal recognition \cite{fernando2003pattern}. In a nutshell, reservoir computing exploits the intrinsic nonlinearity of a physical system to replicate the process of nodal connections in a neural network to extract features from time series signals for machine perception \cite{tanaka2019recent, shougat2021information}. Reservoir computing directly conducts computations in an analog fashion by using the physical system, which largely eliminates the necessity of separate data storage, organization, and machine learning perception. Notably, reservoir computing is naturally suited for audio processing tasks, which are a subset of time series signals. It should also be noted that more traditional, non-physical reservoir computing approaches have seen widespread use for the Internet of Things. Some of these examples include dynamic spectrum management \cite{song2021deep}, network traffic prediction \cite{zhou2022multiscale}, and UAV trajectory design \cite{nasr2022single}.

Researchers have explored many physical systems to operate as reservoir computers for temporal signal processing. These systems include the field-programmable gate array (FPGA) \cite{moran2021hardware}, chemical reactions \cite{usami2021materio}, memristors \cite{moon2019temporal}, superparamagnetic tunnel junctions \cite{mizrahi2018neural}, spintronics \cite{grollier2020neuromorphic}, attenuation of wavelength of lasers in special mediums \cite{larger2017high}, MEMS (microelectromechanical systems) \cite{barazani2020microfabricated}, and others \cite{tanaka2019recent, kan2021simple}. Though these studies have demonstrated that reservoir computing could handle audio signal processing, the physical system for computing is usually very cumbersome \cite{larger2017high}, and they all require preprocessing of the original audio clips using methods such as the Mel spectrum, which largely cancels the benefits of reducing the computational requirements of machine learning via reservoir computing. More importantly, to boost the computational power, conventional reservoir computing techniques use time-delayed feedback achieved by a digital to analog conversion \cite{appeltant2011information}, and the time-delayed feedback will hamper the processing speed of reservoir computing while drastically increasing the envelope of energy consumption for computing. We suggest that the less-than-satisfactory performance of physical reservoir computing is largely caused by the insufficient computational power of the computing systems chosen by the previous works. 

Recently, we have discovered that the Hopf oscillator, which is a common model for many physical processes, has sufficient computational power to conduct machine learning. Although this is a very simple physical system, computing can be achieved without the need of additional data handling, time-delayed feedback, or auxiliary electrical components \cite{shougat2021hopf, shougat2022dynamic,li2021stochastic, li2021four}. Notably, reconfigurable performance can also be achieved for traditional, non-physical reservoir computers \cite{sun2018resinnet}. The performance of the Hopf oscillator reservoir computer on a set of benchmarking tasks (e.g., logical tasks, emulation of time-series signals, and prediction tasks) is exceptional compared to much more complex physical reservoirs. This paper is an extension of previous work to further demonstrate the outstanding capabilities of the Hopf reservoir computer for audio signal recognition tasks. These results point to the efficacy of using this type of reservoir computer for edge computing, which could pave the way to obtaining edge artificial intelligence and decentralized deep learning in the foreseeable future. 


\section{Hopf Oscillator \& Reservoir}\label{sec:hopfRC}

The forced Hopf oscillator is represented in eq. \ref{eq:hopf} \cite{nayfeh2008applied,li2021four}:

\begin{equation}\label{eq:hopf}
\begin{array}{ll}
\dot{x} & = \big(\mu-(x^2+y^2) \big)x - \omega_0 y + A \sin(\Omega t) \\
\dot{y} & = \big(\mu-(x^2+y^2) \big)y + \omega_0 x
\end{array}
\end{equation}

In the above equations, $x$ and $y$ refer to the first and second states of the Hopf oscillator, respectively. The $\omega_0$ term is the resonance frequency of the Hopf oscillator. The $\mu$ parameter affects the radius of the limit cycle motion. For example, without external forcing, the Hopf oscillator would have a limit cycle of radius $\mu$, and it would oscillate at a frequency of $\omega_0$. This parameter also loosely correlates with the quality factor of the oscillator. $A$ is the amplitude of a sinusoidal force.

For the oscillator to classify audio signals, an external forcing signal that contains the audio signal, $a(t)$ is constructed, which is shown in eq. \ref{eq:hopfRc1}; this is then used as input to the Hopf oscillator. The modified Hopf oscillator as a reservoir is represented by eqs. \ref{eq:hopfRc2} and \ref{eq:hopfRc3}:

\begin{align}
f(t) & = 1 + a(t)\label{eq:hopfRc1}\\
\dot{x} & = \big(\mu f(t)-(x^2+y^2) \big)x - \omega_0 y + A f(t) \sin(\Omega t)\label{eq:hopfRc2}\\
\dot{y} & = \big(\mu f(t)-(x^2+y^2) \big)y + \omega_0 x\label{eq:hopfRc3}
\end{align}

\begin{figure}[htpb]
    \centering
    \includegraphics[keepaspectratio=true, width=1\columnwidth]{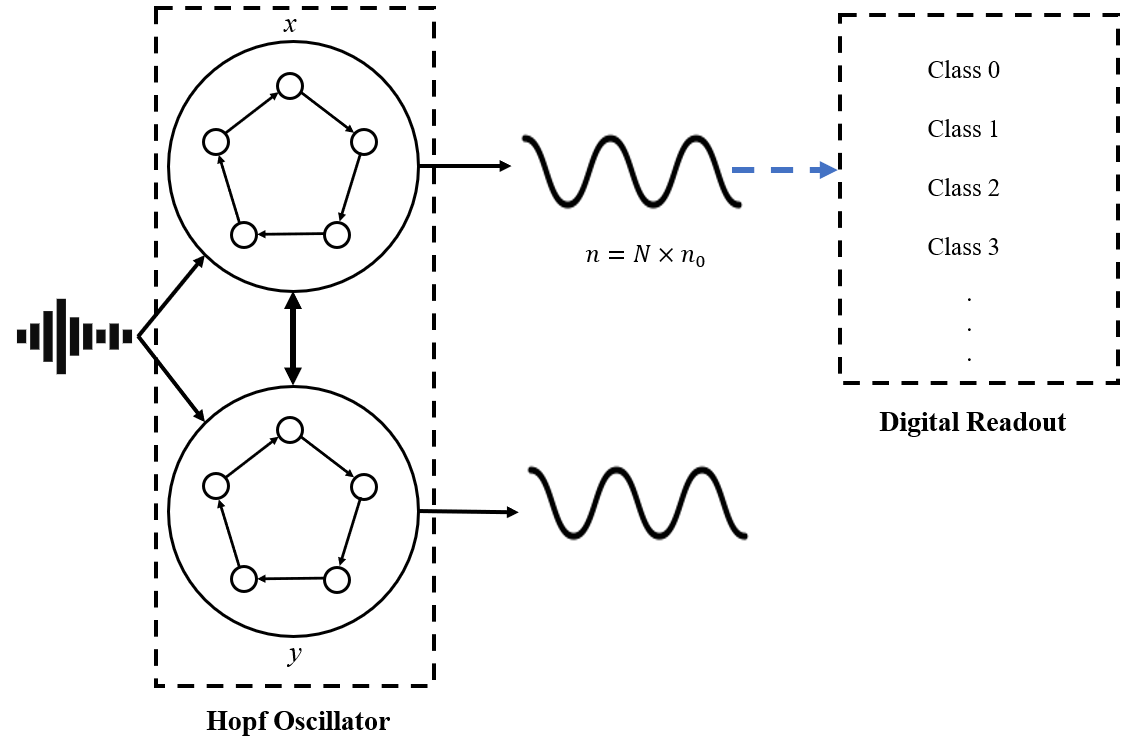}
    \caption{A schematic showing the nodal connections within a Hopf oscillator for reservoir computing. The original signal, $f(t)$, is sent to the two states of the oscillator (i.e., two physical nodes). Each physical node generates $N$ virtual nodes in time series. The digital readout layers (i.e., machine learning algorithm) will read $n$ samples from the node $x$ of the oscillator (note that we only use one node for audio classification in the present paper). $n_0$ corresponds to the number of samples from the original audio signal, and $N$ refers to the number of virtual nodes controlled by the readout mechanisms. The signal from the reservoir is then sent to a neural network, which is indicated by the blue dashed arrow; this neural network is described in Fig. \ref{fig:machineLearningArchitecture}. The digital readout will classify the $n$ samples corresponding to one audio clip to its class. }
    \label{fig:schematic}
\end{figure}

The external signal, $f(t)$, is composed of a DC offset and the audio signal, $a(t)$. The DC offset ensures that the radius parameter is non-negative. This external signal is injected into both the radius parameter, $\mu$, and the sinusoid, $A\sin(\Omega t)$. The Hopf oscillator dynamically responds to the audio signal, and the $x$ state corresponds to the audio features for the machine learning audio classification task. The $y$ state, although not explicitly used in the classification task (as depicted in Fig. \ref{fig:schematic}), likely stores information and aids in the computational task. Unlike the original form of the Hopf oscillator reservoir computer, we use the Hopf oscillations to extract audio features for classification instead of directly using the two state outputs for time series signal prediction \cite{shougat2021hopf}. As such, several changes are made in the computational scheme of the Hopf oscillator reservoir computer. First, this formulation of the reservoir does not include time-multiplexing or a masking procedure. Conventional reservoir computing uses a preset mask multiplying the reservoir outputs to create neurons in the reservoir system. The training of the mask equates to updating parameters when training the digitally realized neural networks. However, this method is memory expensive and inefficient for audio signal processing, since the length of the mask should be sufficient to cover the length of audio clip and the nodal connections necessary for the signal classification. Instead of training masks, we use a more efficient multiple layer convolutional neural network readout to directly feed forward the reservoir outputs and train the connections between each layer as the parameters. Second, no Gaussian noise is added, as the audio signals already have background noise. Third, instead of using a pseudo-period to guide the training of the machine learning readout, we use the number of samples collected for classification to control the nodal connections within each collected feature point generated from the reservoir processing 1D audio data. $N$ virtual nodes means that for each sampling point of the original audio, the reservoir will generate $N-1$ nodal connections in 1D for each reservoir state for classification. For example, with $N$ virtual nodes, a sampled audio data point is processed by the physical node (i.e., $x$ in Fig. \ref{fig:schematic}) $N-1$ times, which creates $N$ feature points from one audio sample and $N-1$ nodal connections in these $N$ feature points. In the current paper, we set $N$ to 100 for audio processing. This method hinders the sampling speed of the audio signals. Thus, we resample the original full resolution audio data to ensure that we operate experiments within a relatively short period of time. It is worth noting that the length of the audio clips for each classification event effectively builds the pseudo-period in the traditional context of the reservoir computing via time-delayed feedback loops (i.e., a fixed length of the audio will produce one classification result with details provided later). The eventual nodal connection of the Hopf reservoir computer and output handling could be conceptualized as Fig. \ref{fig:schematic}.

\begin{figure}[htpb]
    \centering
    \includegraphics[keepaspectratio=true, width=1\columnwidth]{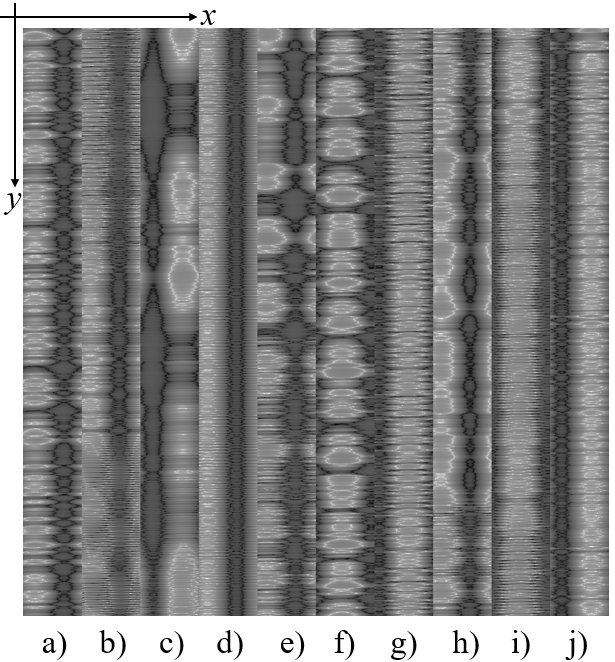}
    \caption{Sample feature maps generated by the Hopf oscillator corresponding to different audio events. Each audio clip has a length of 1 sec sampled at 4000 Hz. The $x$-axis follows the arithmetic order of the virtual nodes, and the $y$-axis is the time. The reservoir is set to have 100 nodes for the test. The grayscale value (from 0 to 1) of each pixel corresponds to the signal strength of each data point (i.e., feature point of the audio signal). a) Air conditioner. b) Car horn. c) Children playing. d) Dog barking. e) Drilling. f) Engine idling. g) Gunshot. h) Jackhammer. i) Siren. j) Street music.}
    \label{fig:featureMaps}
\end{figure}

Here, the Hopf reservoir computer is used to compute feature maps, with several representative examples shown in Fig. \ref{fig:featureMaps}. The $x$-axis follows the numerical order of the virtual nodes, and the $y$-axis is the time. The value of the feature map is rescaled from 0 to 1. Consecutive convolutional layers, followed by the flattened layer and fully-connected layers depicted in Fig. \ref{fig:machineLearningArchitecture}, construct the machine learning readout for processing the audio signal outputs from the reservoir, which is further described in Section \ref{sec:methods}. Note that a similar approach is applied in the SOTA urban sound recognition on edge devices \cite{yun2022infrastructure}, though we eliminate the computationally expensive preprocessing of the Mel spectrogram by offloading feature extraction to the reservoir computer. More important, our approach could use a very coarse sampling (4000 Hz was used here) instead of the Mel spectrogram applied in \cite{yun2022infrastructure} to capture the granularity of the audio signals. A detailed comparison is provided in the subsequent section to demonstrate the superior feature extraction from the Hopf reservoir computer. 

\section{Methods}\label{sec:methods}

\begin{figure}[htpb]
    \centering
    \includegraphics[keepaspectratio=true, width=1\columnwidth]{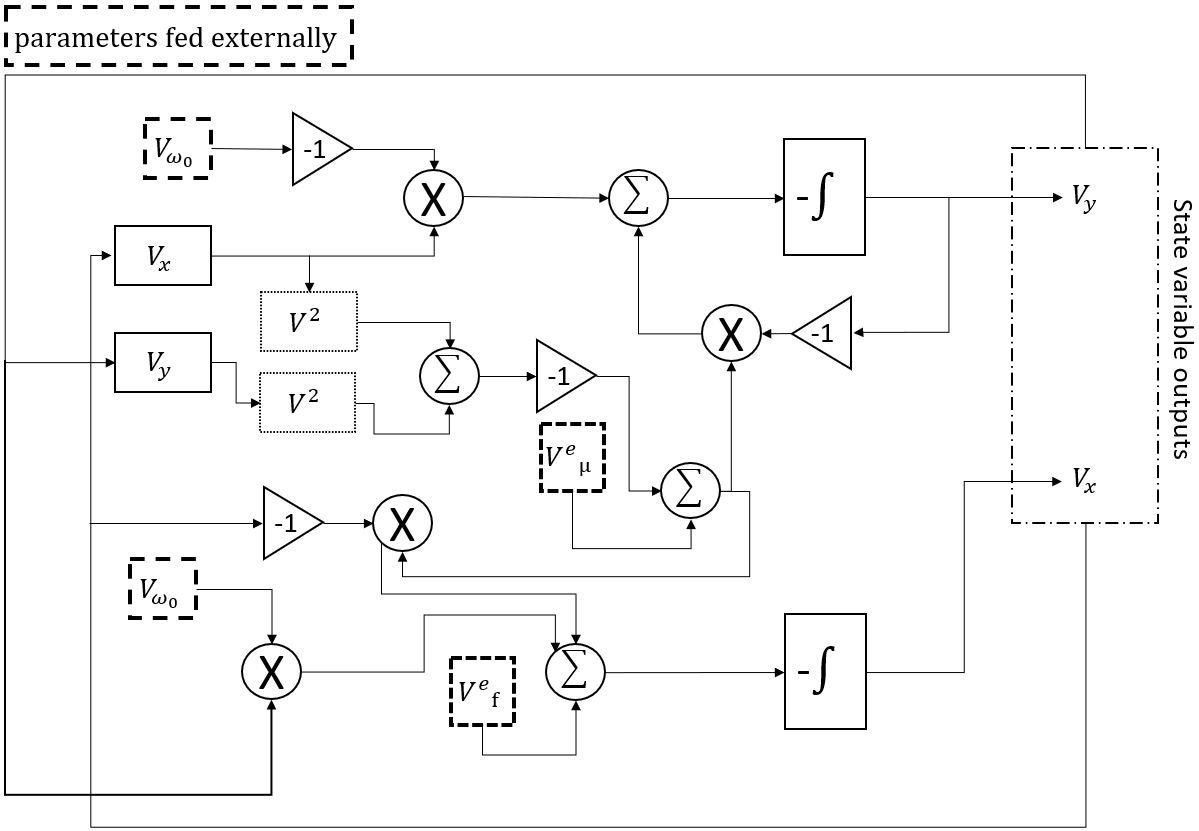}
    \caption{A simplified circuit schematic of the Hopf reservoir computer.}
    \label{fig:schematic_ckt}
\end{figure}

The Hopf physical reservoir computer is realized through a proprietary circuit design proposed by \cite{shougat2021hopf}. Following the schematic given in Fig. \ref{fig:schematic_ckt}, the circuit is implemented using TL082 operational amplifiers and AD633 multipliers. The input audio signal is first normalized to the range from $-1$ to $+1$ and mixed with the sinusoidal forcing signal in MATLAB, then it is sent to the circuit by a National Instrument (NI) cDAQ-9174 data I/O module. The outputs from the circuit, referred to as the $x$ and $y$ states of the Hopf oscillator, are collected with a sampling rate of $10^5$ samples/s by the same NI cDAQ-9174 for later machine learning processing.
  
Three datasets are employed in the sound recognition experiments. These consist of urban sound recognition, Qualcomm voice command, and spoken digits. The urban sound recognition dataset consists of 873 audio clips of 10 classes, which are high quality urban sound clips recorded in New York City \cite{salamon2014dataset}. Each audio clip is four seconds long with a sampling rate of at least 44.1 kHz. Compared to commonly available datasets, we have an extremely small number of samples. 
To demonstrate reconfigurability of the Hopf reservoir computer for audio processing, the Qualcomm voice command dataset is also used. This dataset consists of 4270 audio clips with each clip lasting 1 second, which are four wake words that are collected from speakers with diverse speaking speeds and accents \cite{kim2019query}. From the dataset, we use 1000 clips for experiments. Compared to the previous urban sound recognition case, the only difference in the processing algorithm is the retraining of the output portion (i.e., after convolution layers) of the machine learning readout (details are discussed in the later part of the methodology section and results section of the paper). To compare the proposed Hopf reservoir with other reservoirs, we also conduct an experiment of spoken digits recognition, which serves as the standard benchmarking test for reservoir computing. The spoken digits dataset consists of 3000 audio clips, which are spoken by five different speakers \cite{freeSpokenData}. As with the Qualcomm voice command dataset, the total number of audio clips for the experiments is set to be only 1000.

For the sake of processing speed, we resample each audio clip with a sampling rate of 4000 Hz and normalize the data to the range from $-1$ to $+1$ before sending to the analog circuit. 80\% of the outputs from the circuit are used for training the machine learning model with the remaining 20\% used for testing.

In Fig. \ref{fig:schematic}, the nodal connections of the Hopf physical reservoir computer are shown. Although we only collect a 1D data stream from the Hopf circuit, the data stream consists of both input signals and the response from the virtual nodes defined by the sampling speed of the signals \cite{jacobson2021hybrid}. We follow this principle of arranging and manipulating signals by their virtual nodes. The output from the circuit reservoir is first activated using a inverse hyperbolic tangent function \cite{shougat2021hopf, miller1995activity}

\begin{equation}\label{eq:tanh}
    x_{feature} = {\tanh^{-1}{x}}   
\end{equation}

\begin{figure}[htpb]
    \centering
    \includegraphics[keepaspectratio=true, width=1\columnwidth]{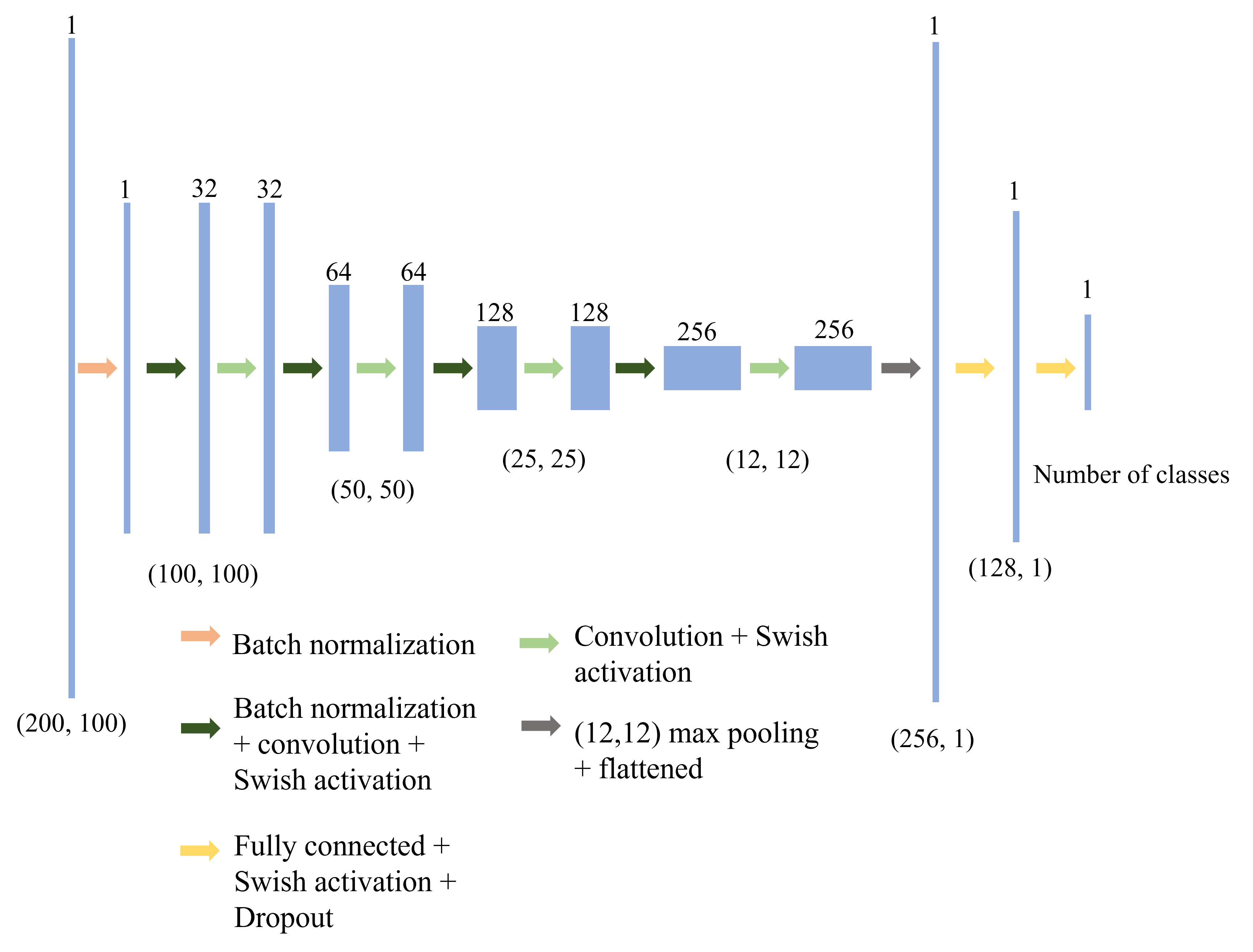}
    \caption{A schematic showing the convolutional neural network-based machine learning readout for classification of the audio events using the Hopf reservoir computer. The light blue boxes in the figure correspond to the feature maps generated from each machine learning operations. The arrows are the different machine learning operations. The numbers above the light blue boxes are the depth of feature maps, and the bottom numbers are the length and the width of the feature maps, respectively. A max pooling with a size of (2,2) is also operated after two consecutive convolutions to reduce the dimension of the feature maps. Note for the length and width, we only label the dimensions that are changed after machine learning operations.}
    \label{fig:machineLearningArchitecture}
\end{figure}

Subsequently, the activated output is rearranged by the order of the virtual nodes as the feature maps for the machine perception. A sample feature map rendering consisting of 10 different classes of urban sound is shown as Fig. \ref{fig:featureMaps}. The Hopf reservoir computer produces this feature map as described in Section \ref{sec:hopfRC}, which is then used as an input to the neural network shown in Figure \ref{fig:machineLearningArchitecture}. Effectively, the Hopf reservoir computer is offloading the costs of the computationally expensive Mel spectrum. A Swish activation \cite{ramachandran2017searching} is employed to boost the performance of the machine learning model on processing sparse neuron activation (i.e., dead neuron problems) and the overall accuracy of the machine learning model processing audio data. Note that a future version of the machine learning software using skipped connection (generating residual networks) \cite{he2016deep} will further boost the robustness of the software for large set of data. Each 1 second clip of the outputs is further skip-sampled to a 200 (number of time samples) $\times$ 100 (number of virtual nodes) for machine learning processing (as labeled in Fig. \ref{fig:machineLearningArchitecture}. The machine learning algorithm is implemented using Keras \cite{keras} with a TensorFlow backend. The training is conducted on an Nvidia RTX 2080Ti GPU and uses an Adam optimizer with the default learning rate of 0.001 \cite{kingma2014adam}. The loss function is cross entropy \cite{de2005tutorial}. The batch size during training is 5; the epochs is 100 for urban sound recognition dataset, 20 for Qualcomm voice command dataset, and 100 for the spoken digits.

\section{Results}\label{sec:results}

\subsection{Results for Urban Sound Recognition Dataset}

\begin{figure}[htpb]
    \centering
    \includegraphics[keepaspectratio=true, width=1\columnwidth]{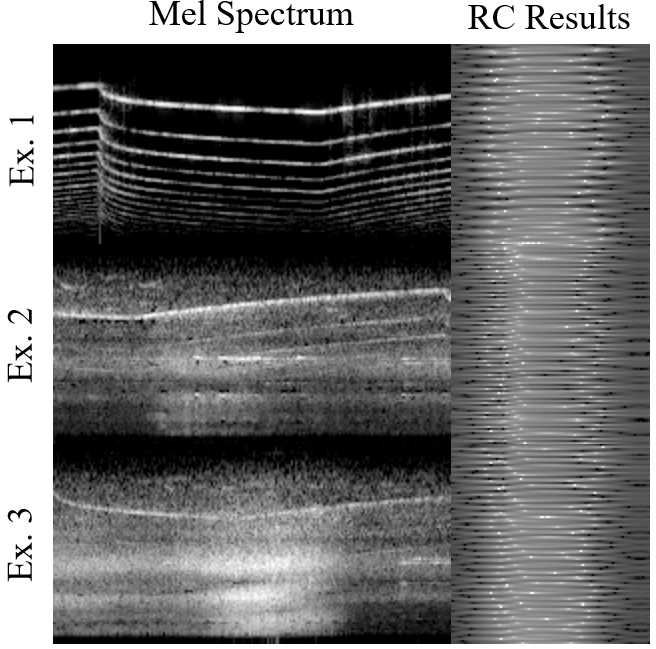}
    \caption{The Mel spectrum is compared with the Hopf RC for the urban sound recognition task. From the top to the bottom, three examples of the siren class are presented. In the left column, the energy of the Mel spectrum is shown. The Mel spectrum operation is conducted upon samples that are four seconds long with a 44.1 kHz sampling rate. The total number of frequency bands is set to be 100, and the time step is set to be 0.025 seconds. In the right column, the audio features extracted from the Hopf reservoir computer for the same samples, such that each 1 second audio clip is downsampled to 4000 Hz and the number of virtual nodes is set to 100.}
    \label{fig:urbanComparison}
\end{figure}

First, we present the results of the Hopf reservoir computer for an urban sound recognition task. As shown in Fig. \ref{fig:urbanComparison} in the left column, the audio features from the Mel spectrum operations (as calculated on the audio clips with a 44.1 kHz sampling rate) show drastic differences between the three examples; using the top example as a reference, the Euclidean distance between the reference and the other two are higher than 25. In comparison, the audio features from the Hopf RC are shown in the right column of Fig. \ref{fig:urbanComparison}; all three examples have a much higher similarity for these three examples (e.g., Euclidean distance $<$12).

\begin{figure}[htpb]
    \centering
    \includegraphics[keepaspectratio=true, width=1\columnwidth]{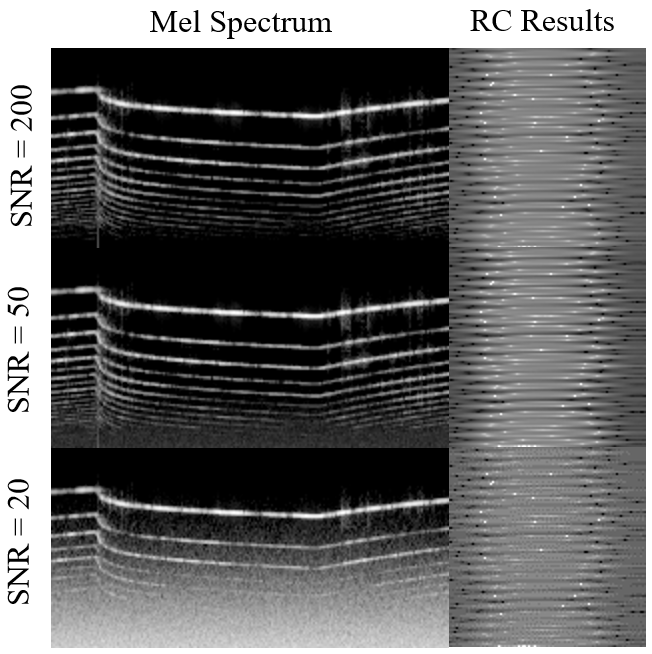}
    \caption{The robustness of the Hopf RC audio extraction is compared with the Mel spectrum for various signal-to-noise ratios (SNRs). For visualization, the siren example shown in the top of Fig. \ref{fig:urbanComparison} is used with different levels of noise. From the top to the bottom, three different amounts of noise were added to the original siren audio example. In the left column, the energy of the Mel spectrum is shown. Note that the result starts to lose low frequency information when the SNR drops to 20. In the right column, the audio features that are extracted using the Hopf RC are shown. Note that the result remains largely the same for all noise levels, even when the SNR is equal to 20.}
    \label{fig:urbanSnrComparison}
\end{figure}

The robustness of the audio classification is also of high importance for real-world applications. To highlight this, the Mel spectrum results are compared with the Hopf RC results for three different noise levels. Using the example in the top row of Fig. \ref{fig:urbanComparison}, white noise is added to the original signal to create different signal-to-noise ratios (SNRs); the audio features of these three new signals are computed with the Mel spectrum (using 44.1 kHz audio sampling rate) and the Hopf reservoir computer (using 4000 Hz audio sampling rate). The output audio features are shown in Fig. \ref{fig:urbanSnrComparison}. It is clearly shown that the Mel spectrum-based audio features lose low frequency information when the SNR is reduced to 20, while the features generated by the Hopf reservoir computer maintain a similar structure with the original audio counterpart, with the Euclidean distance $<$5 for for an SNR of 20. 

\begin{figure}[htpb]
    \centering
    \includegraphics[keepaspectratio=true, width=1\columnwidth]{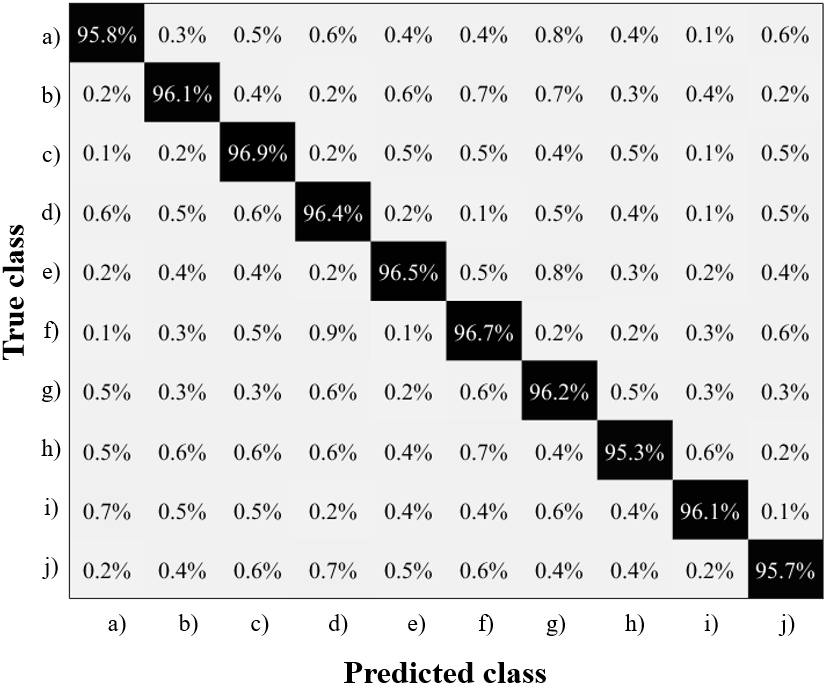}
    \caption{For the urban sound recognition task, the confusion matrix is presented with the recognition accuracy labeled for the ten different audio events. Note that the class labels in this figure are the same as the class labels of Fig. \ref{fig:featureMaps}.}
    \label{fig:urbanConfusion}
\end{figure}

The confusion matrix for the urban sound recognition task is shown in Fig. \ref{fig:urbanConfusion}. The proposed audio recognition approach based on the Hopf reservoir computer has a 96.2\% accuracy. This accounts for a 10\% accuracy improvement compared to \cite{yun2022infrastructure}, with a reduction of $>$94\% of the FLOPS (floating operations per second) for high sampling rate readout and Mel spectrum computation and $\sim{90\%}$ of the audio pieces for training.

\subsection{Results for Qualcomm Voice Command Dataset}

\begin{figure}[htpb]
    \centering
    \includegraphics[keepaspectratio=true, width=1\columnwidth]{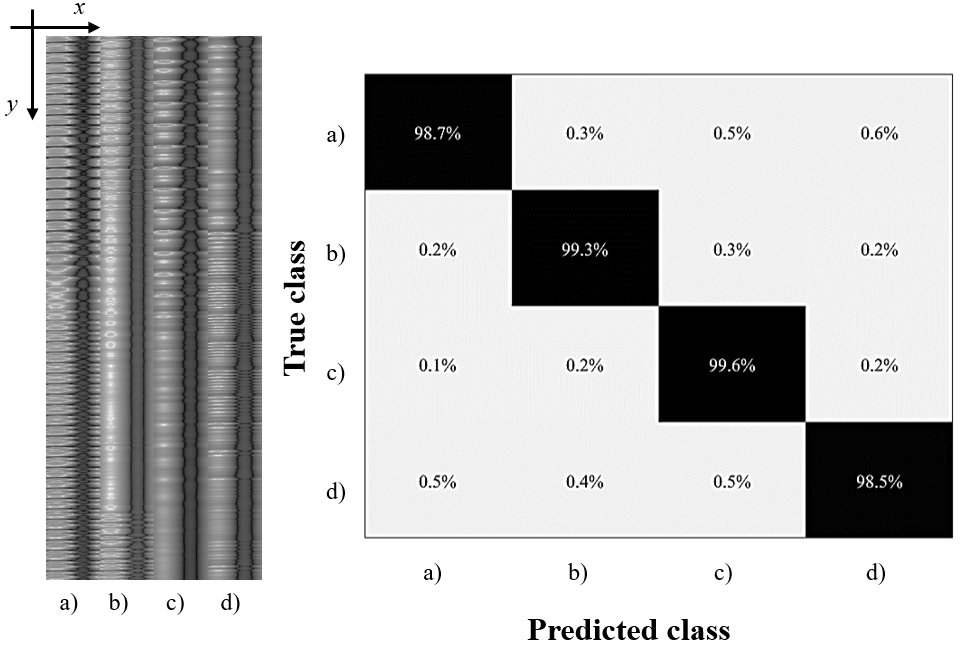}
    \caption{Summary of the results of the Hopf reservoir computer for the Qualcomm voice command task. Left: Examples of the feature maps of different wake words generated by the Hopf reservoir computer. Right: The confusion matrix of the proposed sound recognition system processing Qualcomm wake words. Each label corresponds to: a) ``Hi, Galaxy'', b) ``Hi, Lumia'', c) ``Hi, Snapdragon'', and d) ``Hi, Android''.}
    \label{fig:wakeWords}
\end{figure}

Using the machine learning model trained from the previous test case (i.e., the urban sound recognition task) as the baseline, we test the Qualcomm voice command dataset to demonstrate the reconfigurability of the Hopf reservoir computer audio recognition system. In this experiment, we purposefully reduce the number of epochs to 20 and freeze the CNN portion of the machine learning model to reconfigure the process of the audio recognition system from the urban sound detection task to a voice command task. In the left portion of Fig. \ref{fig:wakeWords}, representative audio features of the four classes are shown, which have significant differences compared to the features of the urban sound events (Fig. \ref{fig:featureMaps}). The audio recognition yields a $>$99\% accuracy, with the confusion matrix depicted in the right portion of Fig. \ref{fig:wakeWords}. Note that the number of parameters trained for this experiment is about 35,000, which accounts for about 300 KB dynamic memory for 8-bit input with a batch size of 5 \cite{gao2020estimating, lin2022device}, demonstrating the feasibility of running the training of the machine learning readout on low-level edge devices consuming Li-Po battery level of power.

\subsection{Results for Spoken Digit Dataset}

\begin{figure}[htpb]
    \centering
    \includegraphics[keepaspectratio=true, width=1\columnwidth]{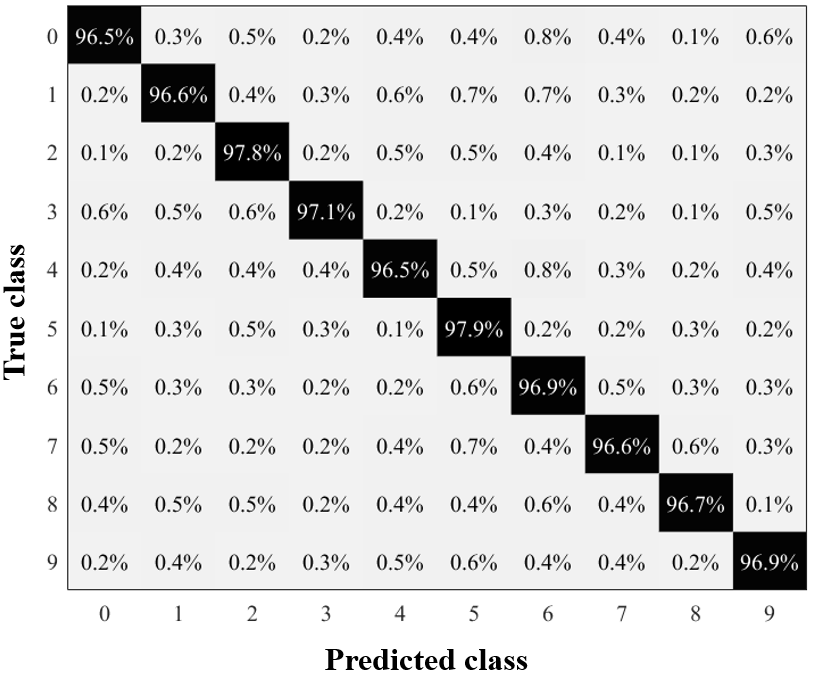}
    \caption{Summary of the results of Hopf reservoir computer conducts spoken digits recognition task. The confusion matrix of the proposed sound recognition system processing  spoken digits dataset with original activation strength and inverse hyperbolic tangent before machine learning readouts.}
    \label{fig:spokenDigit}
\end{figure}

The spoken digit dataset is used to compare the performance of the Hopf reservoir computer for audio recognition with other reservoirs (e.g., \cite{moran2021hardware, usami2021materio, moon2019temporal, mizrahi2018neural, grollier2020neuromorphic, larger2017high, barazani2020microfabricated, kan2021simple}.). As shown in Fig. \ref{fig:spokenDigit}, the Hopf reservoir computer produces an approximately 97\% accuracy for the spoken digit classification task. This result retains the state-of-the-art recognition accuracy on this dataset while only using one physical device (i.e., one consolidated analog circuit) and two physical nodes ($x$ and $y$ states). As a comparison, the best performing reservoir \cite{moon2019temporal} employed 10 memristors and preprocessing of the original audio clips to yield a similar accuracy. We suggest that the vibratory nature of our reservoir largely contributes to the simplicity of the proposed sound event detection system, and the activation of the reservoir using sinusoidal signals boosts the feature extraction of the audio signal using Hopf oscillations (details described later).

\begin{figure}[htpb]
    \centering
    \includegraphics[keepaspectratio=true, width=1\columnwidth]{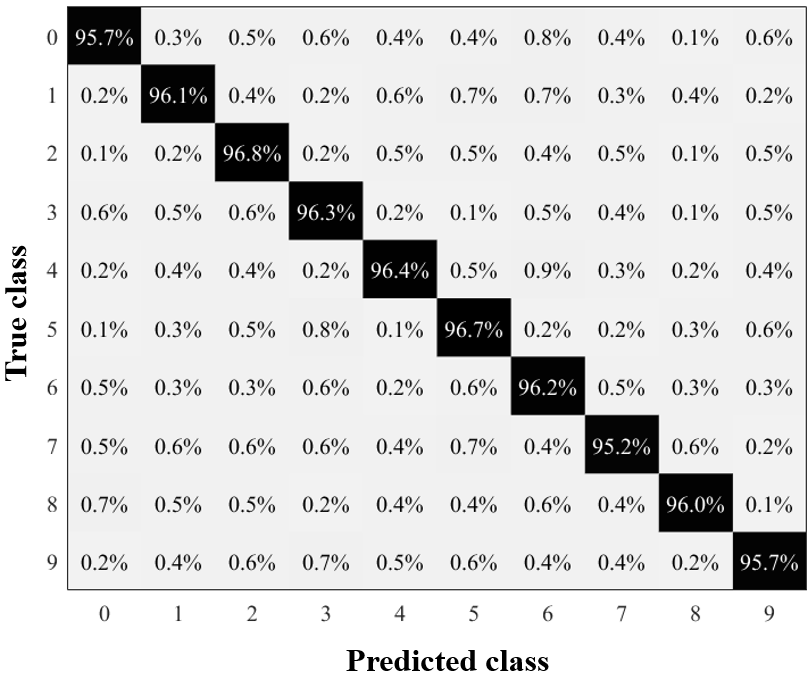}
    \caption{Summary of the results of Hopf reservoir computer conducts spoken digits recognition task. The confusion matrix of the proposed sound recognition system processing  spoken digits dataset with 10 times increase of activation strength and without inverse hyperbolic tangent before machine learning readouts.}
    \label{fig:spokenDigitMod}
\end{figure}

Further, we increase the strength of the activation signal (term $A$ in eq. \ref{eq:hopf}) and discard the inverse hyperbolic tangent activation (eq. \ref{eq:tanh}) before the machine learning readout. The yielded results, which are shown in Fig. \ref{fig:spokenDigitMod}, have a 96\% accuracy compared to the case using eq. \ref{eq:tanh} before sending the $x$ state to the machine learning readout. This suggests that Hopf reservoir computer can be reconfigured not only by its digital readout, similar the traditional physical reservoir computers, but that the Hopf oscillator's computational power could also be drastically enhanced by changing the oscillator's internal physical conditions.

\section{Conclusions}

\subsection{Summary of Results}

In this paper, we present the results of sound signal recognition using reservoir computing technology consisting of a Hopf oscillator \cite{shougat2021hopf, shougat2022dynamic}. Instead of employing computationally expensive preprocessing (e.g., Mel spectrum) commonly used in other studies \cite{larger2017high, moon2019temporal, moran2021hardware, yun2022infrastructure}, we directly take the outputs from the Hopf circuit to process the normalized audio signal for machine learning recognition. We anticipate that this Hopf reservoir computing can be directly implemented to microphones to achieve a future processing-on-the-sensor.

In Section \ref{sec:results}, we systematically demonstrate that our Hopf reservoir computing approach yields a 10\% accuracy improvement on a diverse 10-class urban sound recognition compared to the state-of-the-art results using edge devices \cite{yun2022infrastructure}, whereas we use a surprisingly simple preprocessing by just normalizing the original signal. The wake words recognition results in $>$99\% accuracy using the exact readout machine learning algorithm by only retraining the MLP. This implies that the Hopf reservoir computer will enable inference and reconfiguration on the edge for the sound recognition system. Additionally, compared to other reservoir computing systems (e.g., \cite{moon2019temporal, usami2021materio, kan2021simple, moran2021hardware}), the spoken digit dataset yields superior performance without the need of using complex preprocessing, multiple physical devices, and time-multiplexing; in addition, we have also conducted our benchmarking experiments on far more realistic datasets (i.e., the 10-class urban sound recognition dataset and the 4-class wake words dataset). We demonstrate boosted performance of audio signal processing by changing the activation signal strength of the Hopf oscillator, which implies that there are more degrees of freedom for reconfiguring physical reservoir computers as compared to other reservoir implementations.

Lastly, we carefully crafted the algorithms and preprocessing of the data for sound recognition tasks to keep overall energy consumption, including the digital readout, less than 1 mW based on FLOPS operations and the analog sampling rate. The computational load, which uses less than 700 sound clips of a 10-class dataset for training machine learning models, is well below the envelope of the computational resources possessed by consumer electronic devices. As such, the sound recognition devices using a Hopf reservoir computer could have an effortless integration with devices with untraceable computational load increases.

\subsection{Analysis on the Physical Mechanisms of the Hopf Reservoir Computer Sound Recognition}

Three elements play important roles in the audio signal recognition. The limit cycle system creates an oscillation signal in the temporal domain with a sinusoidal form, which continuously convolves with the incoming audio signal. This effectively creates a 1D short time Fourier transform, generating unique patterns for audio recognition (e.g., Fig. \ref{fig:featureMaps}). Interestingly, this process largely replicates the process of the cochlea in extracting the sound signal features perceptible by the neurons. The nonlinear oscillation of the Hopf oscillator in the temporal direction creates nodal connections of the reservoir computer, corresponding to the neuron connections in DNN. Additionally, the nonlinearity of the Hopf oscillator causes it to respond differently to signals possessing various characteristic features of the audio in a broadband fashion, which produces clean separation of features (Fig. \ref{fig:featureMaps} and Fig. \ref{fig:wakeWords}a).It is worth noting that some recent studies \cite{lenk2018active, PhysRevLett} have demonstrated that the cochlea and its directly-connected neurons creates a limit cycle system using the previous audio signals as activation to dynamically enhance the performance of the cochlea in performing audio signal feature extraction. The physical model of the inner ear can be modeled as a Hopf oscillator with a time-delayed feedback loop using the signals from previous time instants to activate the limit cycle oscillations. The audio signal recognition actually happens in the inner ear instead of in the brain. An interesting future extension of this work is to explore different activation signals to create an artificial ear, which is capable of on-membrane audio recognition. In the meantime, the two states of the Hopf oscillator affect each other with a time delay, which enhances the memory effects essential to the time series signal processing. 

\subsection{Discussion and Future Work}

\begin{figure}[htpb]
    \centering
    \includegraphics[keepaspectratio=true, width=1\columnwidth]{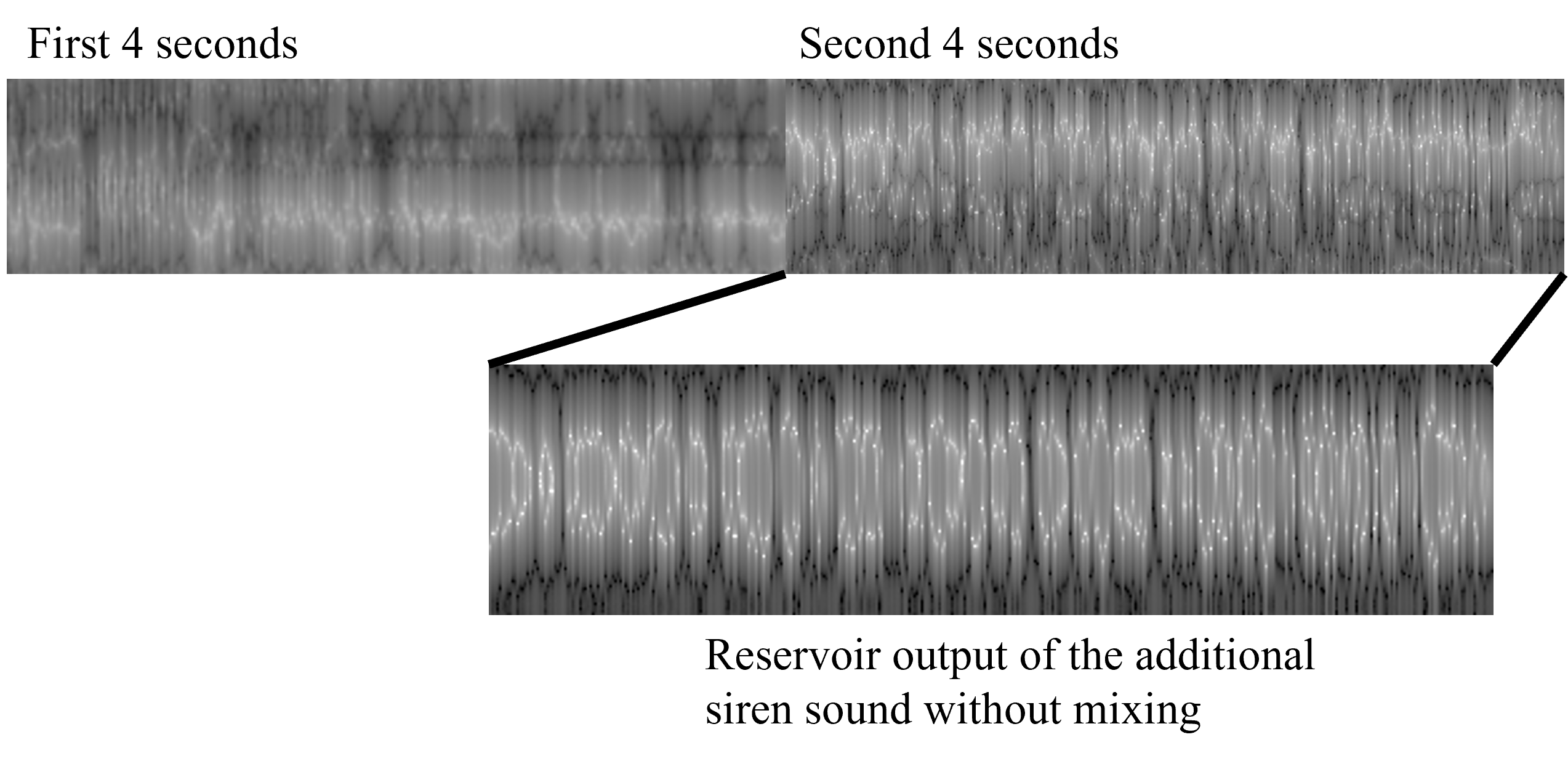}
    \caption{A noise resistance test using audio features generated from the urban sound recognition task. During the first four seconds of this eight second clip, drilling and car horn sounds are mixed, and the last four seconds contains the siren sound with a high amplitude (two times larger as compared to other two audio classes) is added to the mixed data. As shown in the figure, the latter four seconds of audio features shows high similarity compared to the reference siren sound.}
    \label{fig:noiseResistance}
\end{figure}

The unique advantages of the Hopf reservoir computer demonstrated in this paper pave the way for the next generation of smart IoT devices that exploit the unused computational power in sensor networks. Specifically, the physical mechanisms backing reservoir computing also happen in the microphone membrane with carefully crafted activation signals \cite{lenk2018active}. One could imagine that future microphones directly operate sound signal recognition using sensor mechanisms instead of dedicated processing rigs. In addition, as shown in Fig. \ref{fig:featureMaps}, the feature map of sound signals consists of unique patterns that are recognized by a convolutional neural network commonly used for visual signal processing. An extension of the present work will explore the correlations of audio signal feature maps, visual signal feature maps, and other types of time-series data features. As such, reservoir computing could be used as a backbone for multi-modal machine learning in smart IoT paradigms, including sensor fusion, audio video signal combination, and decentralized machine learning. The extremely small amount of training data required for the machine learning operation and clear feature separation described in Section \ref{sec:results} could offer surprisingly satisfactory results, which is essential for many use cases without the luxury of unlimited sizes of datasets (e.g., soft user identification) or with noisy environments (e.g., a mix of different signals). One example is shown in Fig. \ref{fig:noiseResistance}: a eight-second long audio signal consisting of multiple different (i.e., car horn, drilling, and siren) is used to demonstrate the proof-of-concept of Hopf reservoir computer on mixed signal processing. The first four seconds of the audio clip only have car horn and drilling sound. For the last four seconds, the siren sound is added with a higher amplitude. As shown in the figure, the audio features generated from the Hopf reservoir computer has a clearly dominant class on the second half of the data and exhibits visually high correlation with the audio features generated by a clean siren sound with the same Hopf reservoir computer (an Euclidean distance less than 8). We anticipate a pattern matching algorithm originating from computer vision applications could be employed in this type of audio event separation and processing.

There are still limits in the reservoir computing method using the Hopf oscillator in its current form. First, the high accuracy sound event recognition requires many virtual nodes to generate diverse features for machine perception. However, increasing the virtual nodes leads to exponential growth of the sampling rate to read high quality audio data. We are actively seeking solutions to separate audio features from the original signal for recognition and recording, which could decrease the required sampling rate. Second, the current circuit-based physical reservoir separates the process of signal mixing and activation of the circuit. Redesigning the circuit is necessary to simplify signal reading for future system deployment. However, the ultimate version of the Hopf reservoir using MEMS will solve this problem, since the computing will happen on the audio sensing mechanisms. Lastly, the signal processing still relies on a digital readout. Though the algorithm is remarkably simple, a microcontroller unit is needed. We anticipate that the short-term solution will be deploying the optimized machine learning model as firmware (consuming less than 1 MB size of static memory without optimization and less than 256 KB dynamic memory for training upgraded machine learning models). A future goal should be using an analog circuit that could detect the spike signals for audio recognition (similar to neurons) to achieve a fully analog computer on edge devices \cite{ma2019binary}.  


%

\section*{Acknowledgment}

The authors also greatly appreciate the fruitful discussion of the experimental procedures and results with Dr. Omar Zahr and Dr. Helge Seetzen.

\ifCLASSOPTIONcaptionsoff
  \newpage
\fi



\bibliographystyle{IEEEtran}
\bibliography{bibliography}
\end{document}